\documentclass{article}

\usepackage{times}
\usepackage{graphicx} % more modern
\usepackage{subcaption}

\newcommand{\ttt}[1]{\texttt{\small #1}}

\usepackage{natbib}

\usepackage{algorithm}
\usepackage{algorithmic}

\usepackage{hyperref}

\usepackage{amsmath}
\usepackage{amssymb}
\usepackage{ifthen}
\usepackage[utf8]{inputenc}
\usepackage{url}

\usepackage[accepted]{icml2017}

\newcommand\theTitle{Developing Bug-Free Machine Learning Systems With Formal Mathematics}

\icmltitlerunning{\theTitle}

\newcommand\pl[1]{}
\newcommand\dd[1]{}
\newcommand\dhs[1]{}

\usepackage{color}
\definecolor{keywordcolor}{rgb}{0.7, 0.1, 0.1}   % red
\definecolor{tacticcolor}{rgb}{0.0, 0.1, 0.6}    % blue
\definecolor{commentcolor}{rgb}{0.4, 0.4, 0.4}   % grey
\definecolor{symbolcolor}{rgb}{0.0, 0.1, 0.6}    % blue
\definecolor{sortcolor}{rgb}{0.1, 0.5, 0.1}      % green
\definecolor{attributecolor}{rgb}{0.7, 0.1, 0.1} % red

\usepackage{listings}

\usepackage{tikz}

\usetikzlibrary{arrows,positioning}

\tikzstyle{param}=[]
\tikzstyle{det}=[rectangle, draw=black]
\tikzstyle{rand}=[rectangle, rounded corners, draw=blue]

\tikzset{
    >=stealth',
    punkt/.style={
      draw=black,
           text width=5em,
           minimum height=1.5em,
           text centered
    },
    pil/.style={
           ->,
    }
}

\begin{document}

\twocolumn[
\icmltitle{\theTitle}

% It is OKAY to include author information, even for blind
% submissions: the style file will automatically remove it for you
% unless you've provided the [accepted] option to the icml2016
% package.

\begin{icmlauthorlist}
\icmlauthor{Daniel Selsam}{stanford}
\icmlauthor{Percy Liang}{stanford}
\icmlauthor{David L. Dill}{stanford}
\end{icmlauthorlist}

\icmlaffiliation{stanford}{Stanford University, Stanford, CA}

\icmlcorrespondingauthor{Daniel Selsam}{dselsam@stanford.edu}

% You may provide any keywords that you
% find helpful for describing your paper; these are used to populate
% the "keywords" metadata in the PDF but will not be shown in the document
\icmlkeywords{Software, Engineering, Verification, Verify, Errors, Bugs, Formal, Math, Mathematics, Certify, Systems}

\vskip 0.3in
]

\printAffiliationsAndNotice{}

\begin{abstract}
Noisy data, non-convex objectives, model misspecification, and
numerical instability can all cause undesired behaviors in machine
learning systems.  As a result, detecting actual implementation errors
can be extremely difficult.  We demonstrate a methodology in which
developers use an interactive proof assistant to both implement their
system and to state a formal theorem defining what it means for their
system to be correct.  The process of proving this theorem
interactively in the proof assistant exposes all implementation errors
since any error in the program would cause the proof to fail.  As a
case study, we implement a new system, Certigrad, for optimizing over
stochastic computation graphs, and we generate a formal ({\it i.e.}
machine-checkable) proof that the gradients sampled by the system are
unbiased estimates of the true mathematical gradients.  We train a
variational autoencoder using Certigrad and find the performance
comparable to training the same model in TensorFlow.
\end{abstract}

\section{Introduction}
\label{sec:intro}

% ML hard to detect errors
Machine learning systems are difficult to engineer for many
fundamental reasons. First and foremost, implementation errors can be
extremely difficult to detect---let alone to localize and
address---since there are many other potential causes of undesired
behavior in a machine learning system. For example, an implementation
error may lead to incorrect gradients and so cause a learning
algorithm to stall, but such a symptom may also be caused by noise in
the training data, a poor choice of model, an unfavorable optimization
landscape, an inadequate search strategy, or numerical instability.
These other issues are so common that it is often assumed that any
undesired behavior is caused by one of them.
As a result, actual
implementation errors can persist indefinitely without
detection.\footnote{Theano~\cite{bergstra2010theano} has been under
  development for almost a decade and yet there is a recent GitHub
  issue (https://github.com/Theano/Theano/issues/4770) reporting a
  model for which the loss continually diverges in the middle of
  training. Only after various experiments and comparing the behavior
  to other systems did the team agree that it is most likely an
  implementation error. As of this writing, neither the cause of this
  error nor the set of models it affects have been determined.}
Errors are even more difficult to detect in stochastic programs, since
some errors may only distort the distributions of random variables and
may require writing custom statistical tests to detect.

\begin{figure}
Standard methodology: test it empirically

\begin{center}
\begin{tikzpicture}[node distance=1cm, auto,]
  %nodes
  \node[punkt] (program) {Program};

  \node[right=0.5cm of program] (ghost) {};
  \node[above=0.6cm of ghost] (debug) {Debug}
  edge[pil, ->, bend right=15pt] (program);

  \node[punkt, right=1.5cm of program] (test) {Test}
  edge[pil, <-] (program)
  edge[pil, ->, bend right=15pt] (debug);
\end{tikzpicture}
\end{center}

Our methodology: verify it mathematically

\begin{center}
\begin{tikzpicture}[node distance=1cm, auto,]
  %nodes
  \node[punkt, color=red] (specify) {Specify};
  \node[punkt, right=1.0cm of specify] (program) {Program}
  edge[pil, <-] (specify);

  \node[right=0.3cm of program] (ghost) {};
  \node[above=0.6cm of ghost] (debug) {Debug}
  edge[pil, ->, bend right=15pt] (program);

  \node[punkt, right=1.0cm of program, color=red] (prove) {Prove}
  edge[pil, <-] (program)
  edge[pil, ->, bend right=15pt] (debug);
\end{tikzpicture}
\end{center}
\caption{A high-level comparison of our methodology with the
  standard methodology for developing machine learning systems.
  Instead of relying on empirical testing to expose implementation
  errors, we first formally \emph{specify} what our system is required
  to do in terms of the underlying mathematics, and then try to
  formally \emph{prove} that our system satisfies its specification.
  The process of proving exposes implementation errors systematically
  and the (program $\to$ prove $\to$ debug) loop eventually terminates
  with a bug-free system and a machine-checkable proof of correctness.}
\label{fig:intro}
\end{figure}

% Also hard to get right without testing
Machine learning systems are also difficult to engineer because it can
require substantial expertise in mathematics ({\it e.g.} linear
algebra, statistics, multivariate analysis, measure theory,
differential geometry, topology) to even understand what a machine
learning algorithm is supposed to do and why it is thought to do it
correctly. Even simple algorithms such as gradient descent can have
intricate justifications, and there can be a large gap between the
mechanics of an implementation---especially a highly-optimized
one---and its intended mathematical semantics.

% Search for bugs mathematically, instead of empirically
In this paper, we demonstrate a practical methodology for building
machine learning systems that addresses these challenges by enabling
developers to find and eliminate implementation errors systematically
without recourse to empirical testing.  Our approach makes use of a
tool called an \emph{interactive proof
  assistant}~\cite{gordon1979edinburgh, gordon1993introduction,
  harrison1996hol, nipkow2002isabelle, owre1992pvs, coq, delean},
which consists of (a) a programming language, (b) a language to state
mathematical theorems, and (c) a set of tools for constructing formal
proofs of such theorems.  Note: we use the term \emph{formal proof} to mean
a proof that is in a formal system and so can be checked by a machine.

In our approach, developers use the theorem language (b) to state a
formal mathematical theorem that defines what it means for their
implementation to be error-free in terms of the underlying mathematics
({\it{e.g.}}  multivariate analysis). Upon implementing the system
using the programming language (a), developers use the proof tools (c)
to construct a formal proof of the theorem stating that their
implementation is correct. The first draft of any implementation will
often have errors, and the process of interactive proving will expose
these errors systematically by yielding impossible proof
obligations.  Once all implementation errors have been fixed, the
developers will be able to complete the formal proof and be certain
that the implementation has no errors with respect to its
specification.  Moreover, the proof assistant can check the formal
proof automatically so no human needs to understand why the proof is
correct in order to trust that it is. Figure~\ref{fig:intro}
illustrates this process.

% Context: verification + formal mathematics
Proving correctness of machine learning systems requires building on
the tools and insights from two distinct fields: program
verification~\cite{leroy2009formal, klein2009sel4,
  chlipala2013bedrock, chen2015using}, which has aimed to prove
properties of computer programs, and formal
mathematics~\cite{rudnicki1992overview, gonthier2008formal,
  gonthier2013machine, hales2015formal}, which has aimed to formally
represent and generate machine-checkable proofs of mathematical
theorems.  Both of these fields make use of interactive proof
assistants, but the tools, libraries and design patterns developed by
the two fields focus on different problems and have remained largely
incompatible. While the methodology we have outlined will be familiar
to the program verification community, and while reasoning formally
about the mathematics that underlies machine learning will be familiar
to the formal mathematics community, proving such sophisticated
mathematical properties of large (stochastic) software systems is a
new goal and poses many new challenges.

To explore these challenges and to demonstrate the practicality of our
approach, we implemented a new machine learning system,
\emph{Certigrad}, for optimizing over stochastic computation
graphs~\cite{schulman2015gradient}.  Stochastic computation graphs
extend the computation graphs that underly systems like
TensorFlow~\cite{tensorflow2015-whitepaper} and
Theano~\cite{bergstra2010theano} by allowing nodes to represent random
variables and by defining the loss function for a graph to be the expected value
of the sum of the leaf nodes over the stochastic choices.
See Figure~\ref{fig:ex_scg} for an example of a stochastic
computation graph. We implement our system in the Lean Theorem
Prover~\cite{delean}, a new interactive proof assistant still under
active development for which the integration of programming and
mathematical reasoning is an ongoing design goal. We formally state
and prove functional correctness for the stochastic backpropagation
algorithm: that the sampled gradients are indeed unbiased estimates of
the gradients of the loss function with respect to the parameters.

\begin{figure}
\begin{tikzpicture}[node distance=1cm, auto,]
  %nodes
  \node[param] (x) {$x$};

  \node[det, right=0.3cm of x] (gemv1) {$*$}
  edge[pil, <-] (x);

  \node[above=0.5cm of gemv1] (W1) {$W_1$}
  edge[pil, ->] (gemv1);

  \node[det, right=0.3cm of gemv1] (splus1) {$\mathrm{softplus}$}
  edge[pil, <-] (gemv1);

  \node[rand, right=0.3cm of splus1] (z) {$\mathcal{N}(\cdot, I)$}
  edge[pil, <-] (splus1);

  \node[det, right=0.3cm of z] (gemv2) {$*$}
  edge[pil, <-] (z);

  \node[above=0.5cm of gemv2] (W2) {$W_2$}
  edge[pil, ->] (gemv2);
  \node[det, right=0.3cm of gemv2] (sig2) {$\mathrm{sigmoid}$}
  edge[pil, <-] (gemv2);

  \node[det, below=0.3cm of sig2] (cost) {$\mathrm{cost}$}
  edge[pil, <-] (sig2)
  edge[pil, bend left=15, <-] (x);
\end{tikzpicture}
\[ \mathcal{L}(W_1, W_2) = \mathbb{E}_{z \sim \mathcal{N}(\mathrm{softplus}(W_1 x), I)} \left[ \mathrm{cost}(x, \sigma(W_2 z)) \right] \]
\caption{An example stochastic computation graph representing a simple
  variational autoencoder. Stochastic nodes are indicated by rounded rectangles.
  The loss function for the graph is the expected value of the cost node
  over the stochastic choices, which
  in this case is a single sample from a Gaussian distribution.}
\label{fig:ex_scg}
\end{figure}

We note that provable correctness need not come at the expense of computational
efficiency: proofs need only be checked once during development and they introduce no
runtime overhead.  Although the algorithms we verify
in this work lack many optimizations, most of the running time when training
machine learning systems is spent multiplying matrices, and we are
able to achieve competitive performance simply by linking with an
optimized library for matrix operations (we used Eigen~\cite{eigenweb}).\footnote{Note
  that the validity of our theorem becomes contingent on Eigen's
  matrix operations being functionally equivalent to the versions we
  formally proved correct.}  To demonstrate practical feasibility empirically, we trained an
Auto-Encoding Variational Bayes (AEVB)
model~\cite{kingma2014variational} on MNIST using
ADAM~\cite{kingma2014adam} and found the performance comparable to
training the same model in TensorFlow.

\pagebreak
We summarize our contributions:

\begin{enumerate}
  \setlength\itemsep{1pt}
  \item We present the first application of formal ({\it i.e.}
    machine-checkable) proof techniques to developing machine learning
    systems.

\item We describe a methodology that can detect implementation errors
  systematically in machine learning systems.

\item We demonstrate that our approach is practical by developing a
  performant implementation of a sophisticated machine learning system
  along with a machine-checkable proof of correctness.
\end{enumerate}

\section{Motivation}
\label{sec:motivation}
When developing machine learning systems, many program optimizations
involve extensive algebraic derivations to put mathematical
expressions in closed-form \pl{make a broader statement:
many program optimizations are based on mathematical equivalences, e.g., closed form,
but the mathematical connection is often undocumented and implicit}.
For example, suppose you want to compute
the following quantity efficiently:

\begin{align}
  \int_x \mathcal{N}(x ; \mu, \mathrm{Diag}(\sigma^2)) \log \mathcal{N}(x ; 0, I_{n \times n}).
\end{align}

You expand the density functions, grind through the algebra by hand
and eventually derive the following closed form expression:
%a closed form solution: % PL: solution connotes correctness, which we don't want

\begin{align}
  -\frac{1}{2} \left[\sum_{i=1}^n \left( \sigma_i^2 - \mu_i^2 \right) + n \log 2 \pi \right]
\end{align}

You implement a procedure to compute this quantity and include it as
part of a larger program, but when you run your first experiment, your
plots are not as encouraging as you
hoped. After ruling out many other possible explanations, you
eventually decide to scrutinize this procedure more closely. You
implement a na\"{i}ve Monte Carlo estimator for the quantity above,
compare it against your procedure on a few random inputs and find that
its estimates are systematically biased. What do you do now?  If you
re-check your algebra carefully, you might notice that the sign of \(
\mu_i^2 \) is wrong, but wouldn't it be easier if the compiler checked
your algebra for you and found the erroneous step? Or better yet, if
it did the algebra for you in the first place and could guarantee the
result was error-free?

\pl{also, people might not appreciate the pervasiveness of the problem
from this one example; they might think it's just localized to math,
and they'll assume that TensorFlow people will get it right,
and they just have to write models;
I guess this is something we talked about - who is the audience?
I think this kind of thing occurs more for ML people developing new
algorithms rather than people just using TensorFlow to solve some applied
problem; might be worth adding one sentence explaining the regime
we're interested in
}

\dhs{Actually, this example comes from AEVB where people needed to
  do this derivation to find their model.}

\section{Background: The Lean Theorem Prover}
\label{sec:background}

To develop software systems with no implementation errors, we need a
way to write computer programs, mathematical theorems, and
mathematical proofs all in the same language. All three capabilities
are provided by the new interactive proof assistant
Lean~\cite{delean}.  Lean is an implementation of a logical system
known as the Calculus of Inductive
Constructions~\cite{coquand1988calculus}, and its design is inspired
by the better-known Coq Proof Assistant~\cite{coq}. Our development
makes use of certain features that are unique to Lean, but most of
what we present is equally applicable to Coq, and to a lesser extent,
other interactive theorem provers such as
Isabelle/HOL~\cite{nipkow2002isabelle}.

To explain and motivate the relevant features of Lean, we will walk
through applying our methodology to a toy problem: writing a program
to compute the gradient of the softplus function. We can write
standard functional programs in Lean, such as softplus:
\begin{lstlisting}
def splus (x : $\mathbb{R}$) : $\mathbb{R}$ := log (1 + exp x)
\end{lstlisting}

We can also represent more abstract operations such as integrals and gradients:
\begin{lstlisting}
$\int$ (f : $\mathbb{R}$ $\to$ $\mathbb{R}$) : $\mathbb{R}$                              $\nabla$ (f : $\mathbb{R}$ $\to$ $\mathbb{R}$) ($\theta$ : $\mathbb{R}$) : $\mathbb{R}$
\end{lstlisting}
Here the intended meaning of \lstinline{$\int$ f} is the integral of
the function \lstinline{f} over all of $\mathbb{R}$, while the
intended meaning of \lstinline{$\nabla$ f $\theta$} is the gradient
({\it{i.e.}} the derivative) of the function \lstinline{f} at the
point $\theta$.
Figure~\ref{fig:intgrad} shows how to represent
common idioms of informal mathematics in our formal representation;
note that whereas some of the informal examples are too ambiguous to interpret
without additional information, the Lean representation is always unambiguous.

\begin{figure}
  \begin{center}
  \begin{tabular}{|l|l|}
    \hline
    Informal & Formal \\
    \hline
    \(\int f(x, y)\, dx\) & $\int$ ($\lambda$ \ttt{x}, \ttt{f} \ttt{x} \ttt{y}) \\
    \(\int f(x, y)\) & ? \\
    $\nabla_\theta f(g(\theta)) |_{\theta_0}$ & $\nabla$ ($\lambda$ $\theta$, \ttt{f} (\ttt{g} $\theta$)) $\theta_0$ \\
    $\nabla f(g(\theta))$ & ? \\
    \hline
  \end{tabular}
  \caption{Translating informal usages of the integral and gradient to
    our formal representation. Note that whereas some of the informal
    examples are too ambiguous to interpret without additional
    information, the Lean representation is always unambiguous.  }
  \label{fig:intgrad}
  \end{center}
\end{figure}

We can represent mathematical theorems in Lean as well.
For example, we can use the following predicate to state that a particular function \( f \) is differentiable at a point \( \theta \):
\begin{lstlisting}
is_diff (f : $\mathbb{R}$ $\to$ $\mathbb{R}$) ($\theta$ : $\mathbb{R}$) : Prop
\end{lstlisting}
The fact that the return type of \lstinline{is_diff} is
\lstinline{Prop} indicates that it is not a computer program to be
executed but rather that it represents a mathematical theorem.

We can also state and assume basic properties about the gradient, such as linearity:
\begin{lstlisting}
$\forall$ (f g : $\mathbb{R}$ $\to$ $\mathbb{R}$) ($\theta$ : $\mathbb{R}$), is_diff f $\theta$ $\wedge$ is_diff g $\theta$ $\to$
  $\nabla$ (f + g) $\theta$ = $\nabla$ f $\theta$ + $\nabla$ g $\theta$
\end{lstlisting}

Returning to our running example, we can state the theorem that a particular function \lstinline{f}
computes the gradient of the softplus function:
\begin{lstlisting}
def gsplus_spec (f : $\mathbb{R}$ $\to$ $\mathbb{R}$) : Prop :=
  $\forall$ x, f x = $\nabla$ splus x
\end{lstlisting}
Suppose we try to
write a program to compute the gradient of the softplus function as
follows:
\begin{lstlisting}
def gsplus (x : $\mathbb{R}$) : $\mathbb{R}$ := 1 / (1 + exp x)
\end{lstlisting}
The application \lstinline{gsplus_spec gsplus} represents the
proposition that our implementation \lstinline{gsplus} is correct,
{\it i.e.} that it indeed computes the gradient of the softplus
function for all inputs.

We can try to formally prove theorems in Lean interactively:
\begin{lstlisting}
theorem gsplus_correct : gsplus_spec gsplus :=
@lean:@ $\vdash$ gsplus_spec gsplus
!user:! expand_def gsplus_spec,
@lean:@ $\vdash$ $\forall$ x, gsplus x = $\nabla$ splus x
!user:! introduce x,
@lean:@ x : $\mathbb{R}$ $\vdash$ gsplus x = $\nabla$ splus x
!user:! expand_defs [gsplus, splus],
@lean:@ x : $\mathbb{R}$ $\vdash$ 1 / (1 + exp x) = $\nabla$ ($\lambda$ x, log (1 + exp x)) x
!user:! simplify_grad,
@lean:@ x : $\mathbb{R}$ $\vdash$ 1 / (1 + exp x) = exp x / (1 + exp x)
\end{lstlisting}

The lines beginning with \lstinline{@lean@} show the current
state of the proof as displayed by Lean, which at any time consists of a collection of
goals of the form \lstinline{assumptions $\vdash$ conclusion}.
Every line beginning with \lstinline{!user!} invokes a \emph{tactic}, which
is a command that modifies the proof state in some way such that Lean
can automatically construct proofs of the original goals given proofs of the new ones.
Here the \lstinline{simplify_grad}
tactic rewrites exhaustively with known gradient rules---in this case
it uses the rules for log, exp, addition,
constants, and the identity function.  The final goal is clearly
not provable, which means we have found an implementation error in
\lstinline{gsplus}. Luckily the goal tells us exactly what
\lstinline{gsplus x} needs to return: \lstinline{gsplus x = exp x / (1 + exp x)}.
Once we fix the implementation of \lstinline{gsplus},
the proof script that failed before now succeeds and generates a machine-checkable
proof that the revised \lstinline{gsplus} is bug-free.
Note that we need not have even attempted to implement \lstinline{gsplus} before starting
the proof, since the process itself revealed what the program needs to
compute. We will revisit this phenomenon in \S\ref{subsec:proof}.

In the process of proving the theorem, Lean constructs a formal proof
certificate that can be automatically verified by a small stand-alone
executable, whose soundness is based on a well-established
meta-theoretic argument embedding the core logic of Lean into set
theory, and whose implementation has been heavily scrutinized by many
developers.  Thus no human needs to be able to understand why a proof
is correct in order to trust that it is.\footnote{This appealing
  property can be lost when an axiom is assumed that is not
  true. We discuss this issue further in \S\ref{subsec:axioms}.}

Although we cannot execute functions such as \lstinline{gsplus}
directly in the core logic of Lean (since a real number is an infinite
object that cannot be stored in a computer), we can execute the
floating-point approximation inside Lean's virtual machine:
\begin{lstlisting}
vm_eval gsplus $\pi$ -- answer: 0.958576
\end{lstlisting}

\pl{The writing style wavers between general ('we can prove theorems in Lean')
and specific ('we fix this bug'), which is awkward;
I'd just specialize to the current situation since it reads better
}

\section{Case Study: Certified Stochastic Computation Graphs}
\label{sec:casestudy}
\emph{Stochastic computation graphs} are directed acyclic graphs in
which each node represents a specific computational operation that may
be deterministic or stochastic~\citep{schulman2015gradient}. The loss
function for a graph is defined to be the expected value of the sum of
the leaf nodes over the stochastic choices.
Figure~\ref{fig:ex_scg} shows the stochastic computation graph for a
simple variational autoencoder.

Using our methodology, we developed a system, Certigrad,
which allows users to construct arbitrary stochastic
computation graphs out of the primitives that we provide. The main
purpose of the system is to take a program describing a stochastic
computation graph and to run a randomized algorithm (stochastic
backpropagation) that, in expectation, provably generates unbiased samples of the gradients of the
loss function with respect to the parameters.

\subsection{Overview of Certigrad}

We now briefly describe the components of Certigrad,
some of which have no analogues in traditional software systems.\footnote{The complete development can be found at {\scriptsize\url{www.github.com/dselsam/certigrad}}.}

\emph{Mathematics libraries}. There is a type that represents tensors
of a particular shape, along with basic functions ({\it e.g.}  $\exp$,
$\log$) and operations ({\it e.g.} the gradient, the integral). There
are assumptions about tensors ({\it e.g.}  gradient rules for $\exp$
and $\log$), and facts that are proved in terms of those assumptions
({\it e.g.} the gradient rule for softplus). There is also a type that
represents probability distributions over vectors of tensors, that can
be reasoned about mathematically and that can also be executed
procedurally using a pseudo-random number generator.

\emph{Implementation}. There is a data structure that represents
stochastic computation graphs, as well as an implementation of
stochastic backpropagation. There are also functions that optimize
stochastic computation graphs in various ways ({\it e.g.} by
integrating out parts of the objective function), as well as basic
utilities for training models ({\it e.g.} stochastic gradient
descent).

\emph{Specification}. There is a collection of theorem statements that
collectively define what it means for the implementation to be correct. For
Certigrad, there is one main theorem that states that the stochastic
backpropagation procedure yields unbiased estimates of the true
mathematical gradients. There are also other theorems that state that
individual graph optimizations are sound.

\emph{Proof}. There are many helper lemmas to decompose the proofs into more
manageable chunks, and there are tactic scripts to generate machine-checkable
proofs for each of the lemmas and theorems appearing in the system.
There are also tactic programs to automate certain types of reasoning,
such as computing gradients or proving that functions are continuous.

\emph{Optimized libraries}. While the stochastic backpropagation
function is written in Lean and proved correct, we execute the
primitive tensor operations with the Eigen library for linear
algebra. There is a small amount of C++ code to wrap Eigen operations
for use inside Lean's virtual machine.
\pl{doesn't emphasize the general story which is that there are
general libraries; the glue code isn't that important}

The rest of this section describes the steps we took to develop
Certigrad, which include sketching the high-level architecture,
designing the mathematics libraries, stating the main correctness
theorem and constructing the formal proof. Though many details are
specific to Certigrad, this case study is designed to illustrate our
methodology and we expect other projects will follow a similar
process. Note: Certigrad supports arbitrarily-shaped tensors,
but doing so introduces more notational complexity than conceptual
difficulty and so we simplify the presentation that follows by
assuming that all values are scalars.

\subsection{Informal specification}

The first step of applying our methodology is to write down informally
what the system is required to do. Suppose \( g \) is a stochastic
computation graph with $n$ nodes and (to simplify the notation) that it only
takes a single parameter $\theta$. Then \( g, \theta \) together
define a distribution over the values at the $n$ nodes (\(X_1, \dotsc,
X_n\)). Let \( \mathrm{cost}(g, X_{1:n}) \) be the function that sums
the values of the leaf nodes.  Our primary goal is to write a
(stochastic) backpropagation algorithm \lstinline{bprop} such that for
any graph \( g \),
\begin{align}
\mathbb{E}_{g, \theta} \left[ \ttt{bprop}(g, \theta, X_{1:n}) \right ]
& =
\nabla_\theta \left( \mathbb{E}_{g, \theta} \left[ \ttt{cost}(g, X_{1:n}) \right] \right)
\label{eqn:informalscg}
\end{align}
While this equation may seem sufficient to communicate the
specification to a human with a mathematical background, more
precision is needed to communicate it to a computer.  The next step is
to formalize the background mathematics, such as real numbers
(tensors) and probability distributions, so that we can state a formal
analogue of Equation~\ref{eqn:informalscg} that the computer can
understand.  Although we believe it will be possible to develop
standard libraries of mathematics that future developers can use
off-the-shelf, we needed to develop the mathematics libraries for
Certigrad from scratch.

\subsection{Designing the mathematics libraries}
\label{subsec:axioms}

Whereas in traditional formal mathematics the goal is to construct
mathematics from first principles~\cite{gonthier2013machine,
  hales2015formal}, we need not concern ourselves with foundational
issues and can simply assume that standard mathematical properties
hold. For example, we can assume that there is a type
\lstinline{$\mathbb{R}$} of real numbers without needing to construct
them ({\it e.g.} from Cauchy sequences), and likewise can assume there
is an integration operator on the reals \lstinline{$\int$ (f : $\mathbb{R}$ $\to$ $\mathbb{R}$) : $\mathbb{R}$}
that satisfies the well-known properties without needing to construct it either ({\it e.g.} from Riemann sums).

Note that axioms must be chosen with great care since even a single
false axiom (perhaps caused by a single missing precondition) can in
principle allow proving any false theorem and so would invalidate the
property that all formal proofs can be trusted without
inspection.\footnote{For example, the seemingly harmless axiom
  \(\forall x, x/x = 1 \) without the precondition \( x \neq 0 \) can
  be used to prove the absurdity (\( 0 = 0 * 1 = 0 * (0 / 0) = (0 * 0)
  / 0 = 0 / 0 = 1 \)). If a system assumes this axiom, then a formal
  proof of correctness could not be trusted without inspection since
  the proof may exploit this contradiction.}  However, there are many
preconditions that appear in mathematical theorems, such as
integrability, that are almost always satisfied in machine learning
contexts and which most developers ignore.  Using axioms that omit
such preconditions will necessarily lead to proving theorems that are
themselves missing the corresponding preconditions, but in practice a
non-adversarial developer is extremely unlikely to accidentally construct vacuous
proofs by exploiting these axioms. For the first draft of our system,
we purposely omitted integrability preconditions in our axioms to
simplify the development. Only later did we make our axioms sound and
propagate the additional preconditions throughout the system so that
we could fully trust our formal proofs.

Despite the convenience of axiomatizing the mathematics, designing the
libraries was still challenging for two reasons. First, there were
many different ways to formally represent the mathematical objects in
question, and we needed to experiment to understand the tradeoffs
between the different representations. Second, we needed to extend
several traditional mathematical concepts to support reasoning about
executable computer programs. The rest of this subsection illustrates
these challenges by considering the problem we faced of designing a
representation of probability distributions for Certigrad.

\emph{Representing probability distributions}. Our challenge is to
devise a sufficiently abstract representation of probability
distributions that satisfies the following desiderata:
we can reason about the probability density functions of continuous random variables,
we have a way to reason about arbitrary deterministic functions applied to random variables,
we can execute a distribution procedurally using a pseudo-random number generator (RNG),
the mathematical and procedural representations of a distribution are guaranteed to correspond, and
the mathematics will be recognizable to somebody familiar with the informal math behind stochastic computation graphs.

We first define types to represent the mathematical and procedural
notions of probability distribution. For mathematics, we define a
\lstinline{Func n} to be a functional that takes a real-valued function on $\mathbb{R}^n$ to a scalar:
\begin{lstlisting}
def Func (n : $\mathbb{N}$) : Type := $\forall$ (f : $\mathbb{R}^n$ $\to$ $\mathbb{R}$), $\mathbb{R}$
\end{lstlisting}
The intended semantics is that if \lstinline{p : Func n} represents a distribution on $\mathbb{R}^n$, then \lstinline{p f}
is the expected value of \( f \) over \( p \), {\it i.e.} \( \mathbb{E}_{x \sim p} [ f(x) ] \).

For sampling, we define an \lstinline{Prog n} to be a procedure
that takes an RNG and returns a vector in $\mathbb{R}^n$ along with an updated
RNG:
\begin{lstlisting}
def Prog (n : $\mathbb{N}$) : Type := RNG $\to$ $\mathbb{R}^n$ $\times$ RNG
\end{lstlisting}

We also assume that there are primitive (continuous) distributions
(\lstinline{PrimDist := Func 1 $\times$ Prog 1}) that consist of a
probability density function and a corresponding sampling procedure.
In principle, we could construct all distributions
from uniform variates, but for expediency, we treat other well-understood
distributions as primitive, such as the Gaussian
(\lstinline{gauss $\mu$ $\sigma$ : PrimDist}).

Finally, we define a type of distributions (\lstinline{Dist n}) that
abstractly represents programs that may mix sampling from primitive
distributions with arbitrary deterministic computations. A \lstinline{Dist n} can be
denoted to a \lstinline{Func n} (with the function \lstinline{E}) to reason
about mathematically, and to an \lstinline{Prog n} (with the function
\lstinline{run}) to execute with an RNG.
\pl{E and run need to be introduced more deliberately;
saying 'the function E' presumes the reader already knows about it,
but the reader's seeing it for the first time, which is jarring;
defining their types would also be helpful}

For readers familiar with functional programming, our construction is
similar to a monad.  We allow three ways of constructing a
\lstinline{Dist n}, corresponding to sampling from a primitive
distribution (\lstinline{sample}), returning a value deterministically
(\lstinline{det}), and composing two distributions (\lstinline{compose}):
\begin{lstlisting}
sample ((pdf, prog) : PrimDist) : Dist 1
det (xs : $\mathbb{R}^n$) : Dist n
compose (d$_1$ : Dist m) (d$_2$ : $\mathbb{R}^m$ $\to$ Dist n) : Dist n
\end{lstlisting}
The mathematical semantics of all three constructors are straightforward:
\begin{lstlisting}
E (sample (pdf, prog)) f = $\int$ ($\lambda$ x, pdf x * f x)
E (det xs) f = f xs
E (compose d$_1$ d$_2$) f = E d$_1$ (λ x, (E (d$_2$ x) f))
\end{lstlisting}
as are the procedural semantics:
\begin{lstlisting}
run (sample (pdf, prog)) rng = prog rng
run (det xs) rng = (xs, rng)
run (compose d$_1$ d$_2$) rng =
  let (x, rng') $:= $ run d$_1$ rng in run (d$_2$ x) rng'
\end{lstlisting}

We have defined \lstinline{E} and \lstinline{run} to correspond; we
consider a stochastic program correct if we can prove the relevant
theorems about its \lstinline{Func} denotation, and we sample from it
by passing an RNG to its \lstinline{Prog} denotation.

\subsection{Formal specification}

With the background mathematics in place, the next step is to write
down the formal specification itself. First, we design types for every
other object and function appearing in the informal description. To
start, we need a type \lstinline{SCG n} to represent stochastic
computation graphs on $n$ nodes, and a function
\lstinline{SCG.to_dist} that takes an \lstinline{SCG n} and a scalar
parameter $\theta$ to a distribution over $n$ real numbers
(\lstinline{Dist n}). We also need a function \lstinline{cost} that
takes a graph and the values at each of its nodes and sums the values
at the leaf nodes. Figure~\ref{fig:typespreview} provides the full
types of all objects that will appear in the specification.

\begin{figure}
\begin{lstlisting}
Dist n : Type
E {n : $\mathbb{N}$} (d : Dist n) (f : $\mathbb{R}^n$ $\to$ $\mathbb{R}$) : $\mathbb{R}$
SCG n : Type
SCG.to_dist {n : $\mathbb{N}$} (g : SCG n) ($\theta$ : $\mathbb{R}$) : Dist n
cost {n : $\mathbb{N}$} (g : SCG n) (xs : $\mathbb{R}^n$) : $\mathbb{R}$
\end{lstlisting}
\caption{ The basic types and functions we will need to formally state
  the specification. \lstinline{Dist n} represents a distribution over
  $\mathbb{R}^n$, \ttt{E} is the expected value function,
  \lstinline{SCG n} represents a computation graph on $n$ nodes,
  \ttt{SCG.to\_dist} is the function that samples from an
  \lstinline{SCG n} and yields a distribution over the values at the
  nodes, and \lstinline{cost} sums the values at the leaf nodes of a
  graph.  Curly braces around an argument indicates that it can be
  inferred from context and need not be passed explicitly.}
\label{fig:typespreview}
\end{figure}

Now we can write down a type-correct analogue of the informal specification presented in Equation~\ref{eqn:informalscg}:

\begin{lstlisting}
def bprop_spec (bprop : $\forall$ {n}, SCG n $\to$ $\mathbb{R}$ $\to$ $\mathbb{R}^n$ $\to$ $\mathbb{R}$) : Prop :=
$\forall$ (n : $\mathbb{N}$) (g : SCG n) ($\theta$ : $\mathbb{R}$),
  E (SCG.to_dist g $\theta$) ($\lambda$ xs, bprop g $\theta$ xs)
  =
  $\nabla$ ($\lambda$ $\theta$, E (SCG.to_dist g $\theta$) ($\lambda$ xs, cost g xs)) $\theta$
\end{lstlisting}
Given the mathematics libraries, implementing the other objects and
functions appearing in the specification such as \lstinline{SCG n} and
\lstinline{SCG.to_dist} is straightforward functional programming.

\subsection{Interactive proof}
\label{subsec:proof}
While conventional wisdom is that one would write their program before
trying to prove it correct, the interactive proof process provides so
much helpful information about what the system needs to do that we
began working on the proof immediately after drafting the
specification.  We split the proof into two steps.  First, we
implemented the simplest possible function that satisfied the
specification (that only computed the gradient for a single parameter
at a time and did not memoize at all) and proved that correct. Second,
we implemented a more performant version (that computed the gradient for
multiple parameters simultaneously using memoization) and proved it
equivalent to the first one.

For the first step, we started with a placeholder implementation that
immediately returned zero and let the interactive proof process guide
the implementation. Whenever the proof seemed to require induction on
a particular data structure, we extended the program to recurse on
that data structure; whenever the proof showed that a branch of the
program needed to return a value with a given expectation, we worked
backwards from that to determine what value to return. Proving the
first step also exposed errors in our specification in the form of
missing preconditions. For the specification to hold, we needed to
make additional assumptions about the graph, {\it e.g.} that the
identifier for each node in the graph is unique, and that each leaf
node is a scalar (\lstinline{WellFormed g}). We also needed to assume
a generalization of the differentiability requirement mentioned in
\citet{schulman2015gradient}, that a subset of the nodes determined by
the structure of the graph must be differentiable no matter the result
of any stochastic choices (\lstinline{GradsExist g $\theta$}).

For the second step, we wrote the memoizing implementation before
starting the proof and used the process of proving to test and debug
it. Although the code for memoizing was simple and short, we still
managed to make two implementation errors, one conceptual and one
syntactic. Luckily the process of proving necessarily exposes all
implementation errors, and in this case made it clear how to fix both
of them.

We completed the main proof of correctness before proving most of the
lemmas that the proof depends on, but the lemmas turned out to be true
(except for a few missing preconditions) and so proving them did not
expose any additional implementation errors.  We also completed the
main proof while our axioms were still unsound (see \S\ref{subsec:axioms}).
When we made our axioms sound and propagated the changes we found that our specification required two additional preconditions:
that all functions that are integrated over in
the theorem statement are indeed integrable (\lstinline{IntegralsExist g $\theta$}),
and that the many preconditions needed for pushing the
gradient over each integral in the expected loss are satisfied
(\lstinline{CanDiffUnderInts g $\theta$}). However, tracking these
additional preconditions did not lead to any changes in our actual
implementation.  Figure~\ref{fig:finalspec} shows the final
specification.

\begin{figure}
\begin{lstlisting}
def bprop_spec (bprop : $\forall$ {n}, SCG n $\to$ $\mathbb{R}$ $\to$ $\mathbb{R}^n$ $\to$ $\mathbb{R}$) : Prop :=
$\forall$ (n : $\mathbb{N}$) (g : SCG n) ($\theta$ : $\mathbb{R}$),
  WellFormed g $\wedge$ GradsExist g $\theta$
  $\wedge$ IntegralsExist g $\theta$ $\wedge$ CanDiffUnderInts g $\theta$ $\to$
    E (SCG.to_dist g $\theta$) ($\lambda$ xs, bprop g $\theta$ xs)
    =
    $\nabla$ ($\lambda$ $\theta$, E (SCG.to_dist g $\theta$) ($\lambda$ xs, cost g xs)) $\theta$
\end{lstlisting}
\caption{The final specification for the simplified problem with only
  scalars (as opposed to tensors) and only a single parameter \(
  \theta \). Our actual system supports arbitrarily-shaped tensors and
  differentiating with respect to multiple parameters at once.}
\label{fig:finalspec}
\end{figure}

\subsection{Optimizations}
\label{subsec:optimizations}
We can also use our methodology to verify optimizations that involve
mathematical reasoning.  When developing machine learning models, one
often starts with an easy-to-understand model that induces a gradient
estimator with unacceptably high variance, and does informal
mathematics by hand to derive a new model that has the same
objective function but that induces a better gradient estimator. In
our approach, the user can write both models and use the process of
interactive proving to %systematically test and PL: avoid word 'test'
confirm that they induce the same objective function. Common
transformations can be written once and proved correct so that users
need only write the first model and the second can be derived
and proved equivalent automatically.

As part of Certigrad, we wrote a program optimization that integrates
out the KL-divergence of the multivariate isotropic Gaussian
distribution and we proved once and for all that the optimization is
sound. We also verified an optimization that \emph{reparameterizes} a
model so that random variables do not depend on parameters (and so
need not be backpropagated through). Specifically, the optimization
replaces a node that samples from \( \mathcal{N}(\mu,
\mathrm{Diag}(\sigma^2)) \) with a graph of three nodes that first
samples from \( \mathcal{N}(0, I_{n \times n}) \) and then scales and
shifts the result according to $\sigma$ and $\mu$ respectively.
We applied these two transformations in sequence to a na\"{i}ve
variational-autoencoder to yield the Auto-Encoding Variational Bayes
(AEVB) estimator~\cite{kingma2014variational}.

\subsection{Verifying backpropagation for specific models}

Even though we proved that \lstinline{bprop} satisfies its formal
specification (\lstinline{bprop_spec}), we cannot be sure that it
will compute the correct gradients for a particular model unless we prove
that the model satisfies the preconditions of the specification.
Although some of the preconditions are technically undecidable, in
practice most machine learning models will satisfy them all for simple
reasons. We wrote a (heuristic) tactic program to prove that specific
models satisfy all the preconditions and used it to verify that
\lstinline{bprop} computes the correct gradients for the AEVB model
derived in \S\ref{subsec:optimizations}.

\subsection{Running the system}
\label{subsec:running}

We have proved that our system is correct in an idealized mathematical
context with infinite-precision real numbers. To actually execute the
system we need to replace all real numbers in the program with
floating-point numbers. Although doing so technically invalidates the
specification and can introduce numerical instability in some cases,
this class of errors is well understood~\cite{higham2002accuracy},
could be ruled out as well in principle~\cite{harrison2006floating,
  boldo2015verified, ramananandro2016unified} and is conceptually
distinct from the algorithmic and mathematical errors that our
methodology is designed to eliminate.  To improve performance, we also
replace all tensors with an optimized tensor library (Eigen).  This
approximation could introduce errors into our system if for whatever
reason the Eigen methods we use are not functionally equivalent to
ones we formally reason about; of course developers could achieve
even higher assurance by verifying their optimized tensor code as
well.

\subsection{Experiments}

\begin{figure}
  \begin{center}
    \begin{subfigure}[b]{0.2\textwidth}
      \includegraphics[width=\textwidth]{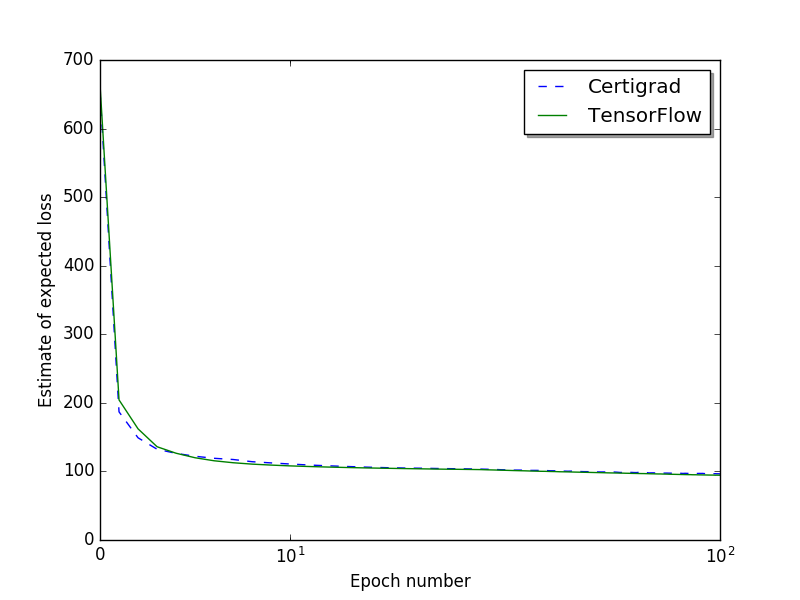}
      \caption{Expected loss vs epoch}
      \label{fig:aevb_loss_vs_epoch}
    \end{subfigure}
    \begin{subfigure}[b]{0.2\textwidth}
      \includegraphics[width=\textwidth]{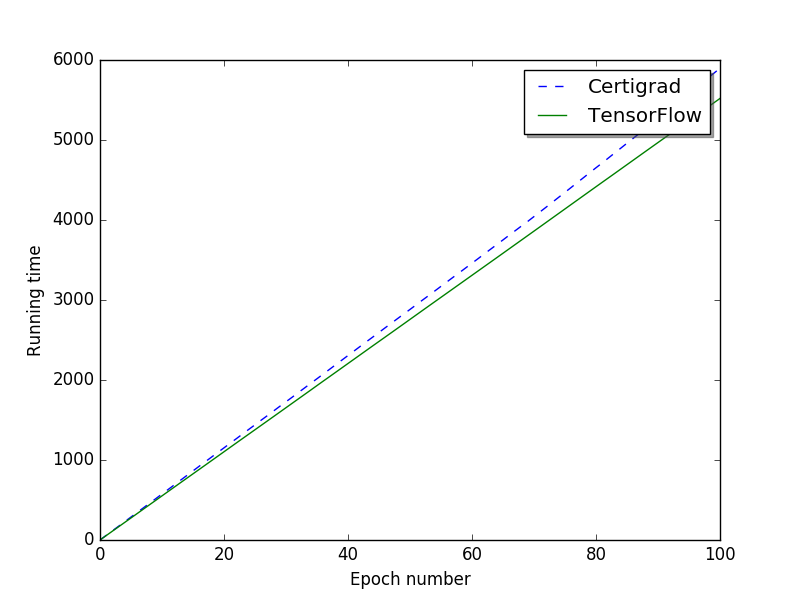}
      \caption{Running time vs epoch}
      \label{fig:aevb_time_vs_epoch}
    \end{subfigure}
\end{center}
\caption{Results of running our certified procedure on an AEVB model,
  compared to TensorFlow. Our system trains just as well and takes only 7\% longer per epoch.}
\label{fig:aevb_plots}
\end{figure}

Certigrad is efficient.  As an experiment, we trained an AEVB model
with a 2-layer encoding network and a 2-layer decoding network on
MNIST using the optimization procedure ADAM~\cite{kingma2014adam}, and
compared both the expected loss and the running time of our system at
each epoch against the same model and optimization procedure in
TensorFlow, both running on 2 CPU cores. We found that the expected
losses decrease at the same rate, and that Certigrad takes only 7\% longer per epoch (Figure~\ref{fig:aevb_plots}).

%% The actual stochastic
%% backpropagation procedure we verified lacks many standard
%% optimizations and so would be much slower than TensorFlow if compared
%% in isolation; however, the computation time is dominated by the tensor
%% operations, which we perform efficiently (as explained in
%% \S\ref{subsec:running}).

\section{Discussion}

Our primary motivation is to develop bug-free machine
learning systems, but our approach may provide significant benefits even
when building systems that need not be perfect.  Perhaps the greatest
burden software developers must bear is needing to fully understand
how and why their system works, and we found that by formally
specifying the system requirements we were able to relegate much of
this burden to the computer.  Not only were we able to synthesize some
fragments of the system (\S\ref{subsec:proof}), we were able to
achieve extremely high confidence that our system was bug-free without
needing to think about how all the pieces of the system fit
together.  In our approach, the computer---not the human---is
responsible for ensuring that all the local properties that the
developer establishes imply that the overall system is
correct. Although using our methodology to develop Certigrad imposed
many new requirements and increased the overall workload
substantially, we found that on the whole it made the development
process less cognitively demanding.

There are many ways that our methodology can be adopted
incrementally. For example, specifications need not cover functional
correctness, not all theorems need to be proved, unsound axioms can be
used that omit certain preconditions, and more traditional code can be
wrapped and axiomatized (as we did with Eigen).  When developing
Certigrad we pursued the ideal of a complete, machine-checkable proof
of functional correctness, and achieved an extremely high level of
confidence that the system was correct. However, we realized many of
the benefits of our methodology---including partial synthesis and
reduced cognitive demand---early in the process before proving most of
the lemmas. Although we could not be certain that we had found all of
the bugs before we made our axioms sound and filled in the gaps in the
formal proofs, in hindsight we had eliminated all bugs early in the
process as well. While a pure version of our methodology may already
be cost-effective for high-assurance applications, we expect that
pragmatic use of our methodology could yield many of the benefits for
relatively little cost and could be useful for developing a wide range
of machine learning systems to varying standards of correctness.

\pagebreak

% PL: don't violate double blind! save for camera ready
\section*{Acknowledgments}
We thank Jacob Steinhardt, Alexander Ratner, Cristina White, William
Hamilton, Nathaniel Thomas, and Vatsal Sharan for providing valuable
feedback on early drafts. We also thank Leonardo de Moura, Tatsu
Hashimoto, and Joseph Helfer for helpful discussions. This work was
supported by Future of Life Institute grant 2016-158712.

\bibliography{main}

\begin{thebibliography}{26}
\providecommand{\natexlab}[1]{#1}
\providecommand{\url}[1]{\texttt{#1}}
\expandafter\ifx\csname urlstyle\endcsname\relax
  \providecommand{\doi}[1]{doi: #1}\else
  \providecommand{\doi}{doi: \begingroup \urlstyle{rm}\Url}\fi

\bibitem[Abadi et~al.(2015)Abadi, Agarwal, Barham, Brevdo, Chen, Citro,
  Corrado, Davis, Dean, Devin, Ghemawat, Goodfellow, Harp, Irving, Isard, Jia,
  Jozefowicz, Kaiser, Kudlur, Levenberg, Man\'{e}, Monga, Moore, Murray, Olah,
  Schuster, Shlens, Steiner, Sutskever, Talwar, Tucker, Vanhoucke, Vasudevan,
  Vi\'{e}gas, Vinyals, Warden, Wattenberg, Wicke, Yu, and
  Zheng]{tensorflow2015-whitepaper}
Abadi, Mart\'{\i}n, Agarwal, Ashish, Barham, Paul, Brevdo, Eugene, Chen,
  Zhifeng, Citro, Craig, Corrado, Greg~S., Davis, Andy, Dean, Jeffrey, Devin,
  Matthieu, Ghemawat, Sanjay, Goodfellow, Ian, Harp, Andrew, Irving, Geoffrey,
  Isard, Michael, Jia, Yangqing, Jozefowicz, Rafal, Kaiser, Lukasz, Kudlur,
  Manjunath, Levenberg, Josh, Man\'{e}, Dan, Monga, Rajat, Moore, Sherry,
  Murray, Derek, Olah, Chris, Schuster, Mike, Shlens, Jonathon, Steiner,
  Benoit, Sutskever, Ilya, Talwar, Kunal, Tucker, Paul, Vanhoucke, Vincent,
  Vasudevan, Vijay, Vi\'{e}gas, Fernanda, Vinyals, Oriol, Warden, Pete,
  Wattenberg, Martin, Wicke, Martin, Yu, Yuan, and Zheng, Xiaoqiang.
\newblock {TensorFlow}: Large-scale machine learning on heterogeneous systems,
  2015.
\newblock URL \url{http://tensorflow.org/}.
\newblock Software available from tensorflow.org.

\bibitem[Bergstra et~al.(2010)Bergstra, Breuleux, Bastien, Lamblin, Pascanu,
  Desjardins, Turian, Warde-Farley, and Bengio]{bergstra2010theano}
Bergstra, J., Breuleux, O., Bastien, F., Lamblin, P., Pascanu, R., Desjardins,
  G., Turian, J., Warde-Farley, D., and Bengio, Y.
\newblock Theano: a {CPU} and {GPU} math expression compiler.
\newblock In \emph{Python for Scientific Computing Conference}, 2010.

\bibitem[Boldo et~al.(2015)Boldo, Jourdan, Leroy, and
  Melquiond]{boldo2015verified}
Boldo, Sylvie, Jourdan, Jacques-Henri, Leroy, Xavier, and Melquiond, Guillaume.
\newblock Verified compilation of floating-point computations.
\newblock \emph{Journal of Automated Reasoning}, 54\penalty0 (2):\penalty0
  135--163, 2015.

\bibitem[Chen et~al.(2015)Chen, Ziegler, Chajed, Chlipala, Kaashoek, and
  Zeldovich]{chen2015using}
Chen, Haogang, Ziegler, Daniel, Chajed, Tej, Chlipala, Adam, Kaashoek, M~Frans,
  and Zeldovich, Nickolai.
\newblock Using crash hoare logic for certifying the fscq file system.
\newblock In \emph{Proceedings of the 25th Symposium on Operating Systems
  Principles}, pp.\  18--37. ACM, 2015.

\bibitem[Chlipala(2013)]{chlipala2013bedrock}
Chlipala, Adam.
\newblock The bedrock structured programming system: Combining generative
  metaprogramming and hoare logic in an extensible program verifier.
\newblock In \emph{ACM SIGPLAN Notices}, volume~48, pp.\  391--402. ACM, 2013.

\bibitem[{{Coq} Development Team}(2015-2016)]{coq}
{{Coq} Development Team}.
\newblock \emph{The {Coq} proof assistant reference manual: Version 8.5}.
\newblock INRIA, 2015-2016.

\bibitem[Coquand \& Huet(1988)Coquand and Huet]{coquand1988calculus}
Coquand, Thierry and Huet, G{\'e}rard.
\newblock The calculus of constructions.
\newblock \emph{Information and computation}, 76\penalty0 (2-3):\penalty0
  95--120, 1988.

\bibitem[de~Moura et~al.(2015)de~Moura, Kong, Avigad, Van~Doorn, and von
  Raumer]{delean}
de~Moura, Leonardo, Kong, Soonho, Avigad, Jeremy, Van~Doorn, Floris, and von
  Raumer, Jakob.
\newblock The {Lean} theorem prover (system description).
\newblock In \emph{Automated Deduction-CADE-25}, pp.\  378--388. Springer,
  2015.

\bibitem[Gonthier(2008)]{gonthier2008formal}
Gonthier, Georges.
\newblock Formal proof--the four-color theorem.
\newblock \emph{Notices of the AMS}, 55\penalty0 (11):\penalty0 1382--1393,
  2008.

\bibitem[Gonthier et~al.(2013)Gonthier, Asperti, Avigad, Bertot, Cohen,
  Garillot, Le~Roux, Mahboubi, O’Connor, Biha, et~al.]{gonthier2013machine}
Gonthier, Georges, Asperti, Andrea, Avigad, Jeremy, Bertot, Yves, Cohen, Cyril,
  Garillot, Fran{\c{c}}ois, Le~Roux, St{\'e}phane, Mahboubi, Assia, O’Connor,
  Russell, Biha, Sidi~Ould, et~al.
\newblock A machine-checked proof of the odd order theorem.
\newblock In \emph{Interactive Theorem Proving}, pp.\  163--179. Springer,
  2013.

\bibitem[Gordon(1979)]{gordon1979edinburgh}
Gordon, Michael~JC.
\newblock Edinburgh lcf: a mechanised logic of computation.
\newblock 1979.

\bibitem[Gordon \& Melham(1993)Gordon and Melham]{gordon1993introduction}
Gordon, Michael~JC and Melham, Tom~F.
\newblock Introduction to hol a theorem proving environment for higher order
  logic.
\newblock 1993.

\bibitem[Guennebaud et~al.(2010)Guennebaud, Jacob, et~al.]{eigenweb}
Guennebaud, Ga\"{e}l, Jacob, Beno\^{i}t, et~al.
\newblock Eigen v3.
\newblock http://eigen.tuxfamily.org, 2010.

\bibitem[Hales et~al.(2015)Hales, Adams, Bauer, Dang, Harrison, Hoang,
  Kaliszyk, Magron, McLaughlin, Nguyen, et~al.]{hales2015formal}
Hales, Thomas, Adams, Mark, Bauer, Gertrud, Dang, Dat~Tat, Harrison, John,
  Hoang, Truong~Le, Kaliszyk, Cezary, Magron, Victor, McLaughlin, Sean, Nguyen,
  Thang~Tat, et~al.
\newblock A formal proof of the kepler conjecture.
\newblock \emph{arXiv preprint arXiv:1501.02155}, 2015.

\bibitem[Harrison(1996)]{harrison1996hol}
Harrison, John.
\newblock Hol light: A tutorial introduction.
\newblock In \emph{International Conference on Formal Methods in Computer-Aided
  Design}, pp.\  265--269. Springer, 1996.

\bibitem[Harrison(2006)]{harrison2006floating}
Harrison, John.
\newblock Floating-point verification using theorem proving.
\newblock In \emph{International School on Formal Methods for the Design of
  Computer, Communication and Software Systems}, pp.\  211--242. Springer,
  2006.

\bibitem[Higham(2002)]{higham2002accuracy}
Higham, Nicholas~J.
\newblock \emph{Accuracy and stability of numerical algorithms}.
\newblock SIAM, 2002.

\bibitem[Kingma \& Welling(2014)Kingma and Welling]{kingma2014variational}
Kingma, D.~P. and Welling, M.
\newblock Auto-encoding variational {B}ayes.
\newblock \emph{arXiv}, 2014.

\bibitem[Kingma \& Ba(2014)Kingma and Ba]{kingma2014adam}
Kingma, Diederik and Ba, Jimmy.
\newblock Adam: A method for stochastic optimization.
\newblock \emph{arXiv preprint arXiv:1412.6980}, 2014.

\bibitem[Klein et~al.(2009)Klein, Elphinstone, Heiser, Andronick, Cock, Derrin,
  Elkaduwe, Engelhardt, Kolanski, Norrish, et~al.]{klein2009sel4}
Klein, Gerwin, Elphinstone, Kevin, Heiser, Gernot, Andronick, June, Cock,
  David, Derrin, Philip, Elkaduwe, Dhammika, Engelhardt, Kai, Kolanski, Rafal,
  Norrish, Michael, et~al.
\newblock sel4: Formal verification of an os kernel.
\newblock In \emph{Proceedings of the ACM SIGOPS 22nd symposium on Operating
  systems principles}, pp.\  207--220. ACM, 2009.

\bibitem[Leroy(2009)]{leroy2009formal}
Leroy, Xavier.
\newblock Formal verification of a realistic compiler.
\newblock \emph{Communications of the ACM}, 52\penalty0 (7):\penalty0 107--115,
  2009.

\bibitem[Nipkow et~al.(2002)Nipkow, Paulson, and Wenzel]{nipkow2002isabelle}
Nipkow, Tobias, Paulson, Lawrence~C, and Wenzel, Markus.
\newblock \emph{{Isabelle/HOL}: a proof assistant for higher-order logic},
  volume 2283.
\newblock Springer, 2002.

\bibitem[Owre et~al.(1992)Owre, Rushby, and Shankar]{owre1992pvs}
Owre, Sam, Rushby, John~M, and Shankar, Natarajan.
\newblock Pvs: A prototype verification system.
\newblock In \emph{Automated Deduction—CADE-11}, pp.\  748--752. Springer,
  1992.

\bibitem[Ramananandro et~al.(2016)Ramananandro, Mountcastle, Meister, and
  Lethin]{ramananandro2016unified}
Ramananandro, Tahina, Mountcastle, Paul, Meister, Beno{\^\i}t, and Lethin,
  Richard.
\newblock A unified coq framework for verifying c programs with floating-point
  computations.
\newblock In \emph{Proceedings of the 5th ACM SIGPLAN Conference on Certified
  Programs and Proofs}, pp.\  15--26. ACM, 2016.

\bibitem[Rudnicki(1992)]{rudnicki1992overview}
Rudnicki, Piotr.
\newblock An overview of the mizar project.
\newblock In \emph{Proceedings of the 1992 Workshop on Types for Proofs and
  Programs}, pp.\  311--330, 1992.

\bibitem[Schulman et~al.(2015)Schulman, Heess, Weber, and
  Abbeel]{schulman2015gradient}
Schulman, John, Heess, Nicolas, Weber, Theophane, and Abbeel, Pieter.
\newblock Gradient estimation using stochastic computation graphs.
\newblock In \emph{Advances in Neural Information Processing Systems}, pp.\
  3528--3536, 2015.

\end{thebibliography}
\bibliographystyle{icml2017}

\end{document}